\documentclass[%
preprint,
amssymb,
aps,prd,
floatfix,
nofootinbib
]{revtex4-1}
\usepackage[centertags]{amsmath}

\usepackage{enumitem}

\newlist{STEP}{enumerate}{1}
\setlist[STEP]{label=\Roman*:}

\usepackage{graphicx}
\usepackage{dcolumn}
\usepackage{bm}
\usepackage{hyperref}
\usepackage{tikz}
\usetikzlibrary{positioning}
\usetikzlibrary{shapes,arrows,arrows,positioning,fit}
\tikzstyle{decision} = [diamond, draw, fill=blue!20, 
  text width=5.5em, text badly centered, inner sep=0pt]
  \tikzstyle{block} = [rectangle, draw, fill=green!20, 
  text width=30em, text centered, rounded corners, minimum height=4em]
  \tikzstyle{line} = [draw, -latex']
  \tikzstyle{cloud} = [draw, ellipse,fill=red!20,
  minimum height=4em]
  \tikzstyle{decisionn} = [diamond, draw, fill=red!20, 
  text width=4em, text badly centered, node distance=4cm, inner sep=0pt]
  \tikzstyle{blockk} = [rectangle, draw, fill=blue!20, 
  text width=8em, text centered, rounded corners, minimum height=2em]
  \tikzstyle{blockkk} = [rectangle, draw, fill=blue!20, 
  text width=8em, text centered, rounded corners, minimum height=2em]
  \def\id{\hat{\mathbb{I}}}

\begin{document}
\title{Extended Uncertainty Principle via Dirac Quantization}

\author{Mytraya Gattu}
\email{mvg6042@psu.edu}
\affiliation{%
 Department of Physics, Indian Institute of Technology Bombay, Mumbai 400076, India
}
\affiliation{%
Department of Physics, 104 Davey Lab, Pennsylvania State University, University Park, Pennsylvania 16802, US}%
\author{S. Shankaranarayanan}%
\email{shanki@phy.iitb.ac.in}
\affiliation{%
 Department of Physics, Indian Institute of Technology Bombay, Mumbai 400076, India
}%

\date{\today}

\begin{abstract}
Unifying quantum theory and gravity remains a fundamental challenge in physics. While most existing literature focuses on the ultraviolet (UV) modifications of quantum theory due to gravity, this work shows that generic infrared (IR) modifications arise when we describe quantum theory in curved spacetime. We explicitly demonstrate that the modifications to the position-momentum algebra are proportional to curvature invariants (such as the Ricci scalar and Kretschmann scalar). Our results, derived through a rigorous application of Dirac's quantization procedure, demonstrate that infrared effects in quantum systems can be axiomatically derived. We study particle dynamics in an arbitrary curved spacetime by embedding them in a higher-dimensional flat geometry.
Our approach, which involves embedding particle dynamics in a higher-dimensional flat geometry and utilizing Dirac's quantization procedure, allows us to capture the dynamics of a particle in 4-dimensional curved spacetime through a modified position-momentum algebra. When applied to various spacetimes, this method reveals that the corrections due to the spacetime curvature are universal. We further compare our results with those derived using extended uncertainty principles. Finally, we discuss the implications of our work for black holes and entanglement.
\end{abstract}

\maketitle

\section{Introduction} \label{sec: 1}

In quantum mechanics \cite{sakurai1995modern,shankar2012principles}, the generators of translation are well defined, as we may linearly superimpose translations in the three orthogonal directions. This is analogous to assuming that the notion of plane waves is always possible~\cite{kempf1996noncommutative} --- a consequence of the fact that the background metric is flat. This naturally leads us to the following question: Is the assumption of plane waves, i.e., precise localization in momentum space, valid in any curved space?~Kempf~\cite{kempf1996path} provided a heuristic answer by demonstrating that the \textcolor{black}{commutation relations} between momentum and position operators is transformed for particles moving in geodesics on a curved geometry. Following this, attempts have been made to identify the relationship between the commutation relations and the curved geometry~\cite{Bolen:2004sq,Park:2007az,Bambi:2007ty,costa2016extended,wagner2022relativistic, petruzziello2021gravitationally}.

However, these attempts are restrictive in two aspects. First, these attempts need the definition of plane waves in the asymptotic limit or, in the case of Kempf~\cite{kempf1996path}, requires the notion of four-momentum to be defined locally everywhere in a curved geometry. Second, many studies are based on a specific geometry, and it is unclear if they can be extended to generic space times. We propose a novel approach to circumvent both these issues in a unified manner. Specifically, we develop a framework that does not require the concept of plane waves in asymptotic time and derive modified commutation relations of globally defined operators for arbitrary curved geometry.

Our approach differs from these approaches in one \emph{fundamental way}: we study the particle dynamics by embedding the 4-dimensional curved space-time in an $(4+m)$-dimensional flat geometry. More specifically, the dynamics of a point particle in a curved space-time $\mathcal{P}$ is mapped to the dynamics in a flat higher-dimensional space $\mathcal{Q}$ with a set of holonomic constraints $\left\{\phi\right\}$. Since the higher-dimensional geometry is flat, the generators of translation in this space are well-defined. Hence, we can superimpose translations in $m$-orthogonal directions linearly. Unlike earlier approaches, embedding in a higher-dimensional flat geometry provides an unambiguous definition of plane waves in arbitrary 4-dimensional space-time.

While the position and momenta satisfy the standard commutation relations (Poisson brackets) in the $(4+m)-$dimensional flat geometry, as we show, the commutation relations of the effective 4-dimensions have corrections related to the curvature of the 4-dimensional space-time. We obtain the commutation relations using Dirac's constraint quantization procedure~\cite{dirac1950generalized}. 

\textcolor{black}{We show that Dirac's procedure allows us to completely capture the dynamics of a particle in 4-dimensional curved space-time through a modified position-momentum algebra.} The modifications are proportional to curvature invariants (like the Ricci and Kretschmann scalar). Thus, the curvature corrections introduce modifications in the largest possible scales, i.e., the \emph{infrared} regime. Interestingly, our analysis provides an independent route to understanding the Extended Uncertainty principle (EUP) that has been proposed in the literature~\cite{Park:2007az,Bolen:2004sq,Bambi:2007ty,costa2016extended}. One of the generic predictions of EUP is the modifications to the standard uncertainty principle:
  \begin{equation} 
  \label{eup}
  \Delta x \Delta p \geq \frac{\hbar}{2}\left(1+\frac{1}{L_{1}^{2}}\Delta x^{2}\right)
  \end{equation}
where ${L_{1}}$ is the characteristic scale of the system.
It can be seen that the modifications to the standard uncertainty principle are dominant at length scales of the order of $L_1$. Given the fact that the Uncertainty principle is related to the commutation relations between non-commuting variables, the EUP \eqref{eup} suggests the following modifications to the commutation relations: 
 \begin{equation}\label{eup-ccr}\left[\hat{x}^{i},\hat{p}_{j}\right] = \imath \hbar \delta^{i}_{j}\left(\hat{1}+\alpha \delta_{kl}\hat{x}^{k}\hat{x}^{l}\right) + \imath \hbar \beta \delta_{j k}\hat{x}^{i}\hat{x}^{k}\end{equation}
where, $\alpha$ and $\beta$ are unknown constants with dimensions of $L^{-2}$.
This work arrives at the above commutation relations from a \emph{first-principles approach}. To our knowledge, such a result has yet to be obtained in the literature.

The paper is organized as follows: In Sec.~\eqref{sec:Procedure}, we {first} discuss the Janet-Cartan theorem and {then} provide detailed steps to obtain the modified position-momentum commutation relations for a massive relativistic particle. In Sec.~\eqref{sec:applications}, we apply the procedure to a few space-times and show that the modifications are proportional to curvature invariants (like Ricci scalar, Kretschmann scalar). 
Finally, in Sec.~\eqref{sec:conc}, we discuss our work's implications and possible future directions.

We use $(+,-,-,-)$ signature for the 4-dimensional space-time metric. We set $c = 1$.

\section{Janet-Cartan theorem and Position-Momentum Algebras}
\label{sec:Procedure}

Viewing our 4-dimensional space-time as a world sheet in a higher-dimensional flat geometry is not new~\cite{Nash1956-mv}. Most commonly, embedding space-time into a higher-dimensional flat space aids in studying its geometry. However, as mentioned in the Introduction, we wish to use the higher-dimensional embedding for a different application.

Our work hinges on the applicability of the Janet-Cartan theorem to any arbitrary four-dimensional space-time~\cite{lidsey1997applications,friedman1965isometric}. Janet-Cartan theorem states that every $(1,n-1)$ dimensional pseudo-Riemannian manifold can be optimally embedded (that is, an embedding of minimum dimension) in a pseudo-Euclidean manifold of dimension no more than $\frac{1}{2}n(n+1)$. Thus, any generic 4-dimensional curved space-time $\mathcal{P}$ can be embedded in a pseudo-Euclidean space $\mathcal{Q}$ of dimension no more than $10$. 

We want to use this geometric property to understand the dynamics of the point-particle. Specifically, we aim to map the dynamics of a particle in a curved space-time $\mathcal{P}$ to its dynamics in a flat space $\mathcal{Q}$ with a set of constraints $\left\{\phi\right\}$. The reason for the mapping rests on the fact that quantum mechanics is both well-defined (for example, the Wiener measure in Path Integrals) and well-understood (for example, the representation of operators is known unambiguously) in Euclidean or pseudo-Euclidean spaces~\cite{wald2009formulation}.

This naturally leads to the following question: How do the space-time symmetries and the curvature affect the quantum dynamics of a point particle?
To understand why this question is relevant, consider two specific cases --- 4-D Minkowski and de Sitter space-times. Minkowski space-time has $SO(1,3)$ symmetry, and the structure of this symmetry group gives rise to the Poincare algebra~\cite{weinberg1995quantum,wigner1939unitary}. However, de Sitter space-time has $SO(1,4)$ symmetry, leading to a different algebra~\cite{moylan1983unitary, thomas1941unitary}. 
From this, we can infer that the space-time symmetries and their curvature 
can affect the quantum dynamics via modifications to the position-momentum algebra. As mentioned in the Introduction, this work aims to derive such modifications for generic 4-D curved space-times.

To this end, let us consider the dynamics of a point particle in a 4-D curved space-time $\mathcal{P}$ such that we can optimally embed $\mathcal{P}$ in a flat-space $\mathcal{Q}$ of dimension $4+m$. We assume that $m$ constraints $\left\{\phi_{i}\right\}_{i=1}^{m}$ that describe $\mathcal{P}$ as embedding are known.

We are, in effect, studying the dynamics of a point particle in a $(4+m)$-D flat space $\mathcal{Q}$ in the presence of $m$-holonomic constraints $\phi_{i}$ on its dynamics. To study the constrained dynamics, we shall follow Dirac's procedure (see Ref.~\cite{dirac1950generalized}).

The free motion of a massive (of mass $m_{0}$) point particle in a flat space $\mathcal{Q}$ (with the metric $\eta_{AB}$) is described by the Lagrangian
\begin{equation}\label{eq:free-particle-lagrangian}
  \mathcal{L}_{0} = -m_{0} \sqrt{-\eta_{AB} \frac{dX^{A}}{d\tau} \frac{dX^{B}}{d\tau}}
\end{equation}
Here, $X^{A}(\tau)$ (with $A$ ranging between $1$ and $4+m$) are the coordinates of the particle as it moves along a trajectory parametrized by $\tau$. \textcolor{black}{Note that the above Lagrangian is a sufficient covariant Lagrangian but is not necessarily the \emph{only} covariant Lagrangian~\cite{2021-Wagner.Guthrie-Arx}. We may choose a Lagrangian such
that the action integral has different values when different parameters are used, for instance: 
\begin{equation}\label{eq:free-particle-lagrangian2}
\mathcal{L} \propto -\eta_{AB} \frac{dX^{A}}{d \sigma} \frac{dX^{B}}{d\sigma}
\end{equation}
Here, $\sigma$ is a scalar related to the affine parameter $\tau$. In this work, we will use the Lagrangian \eqref{eq:free-particle-lagrangian} to do the rest of the analysis. While the analysis using both the Lagrangians are equivalent, in the case of Lagrangian \eqref{eq:free-particle-lagrangian}, we need to fix the reparametrization gauge~\cite{fulop1998reparametrization}.}

\subsection{Total Hamiltonian}

To study the constrained motion of the particle in $\mathcal{Q}$, we enlarge the state space to include $m$ auxiliary variables (i.e. Lagrange multipliers) $\Lambda^{i}$ ($i=1,\dots, m$) and modify the Lagrangian as
\begin{equation}\label{eq:constrained-lagrangian}
 \mathcal{L}_{1} = \mathcal{L}_{0}(X^{A}, \frac{dX^{A}}{d\tau}) + \Lambda^{i} \phi_{i}(X^{A})
\end{equation}

The first step in Dirac's procedure is to construct the total Hamiltonian after identifying any primary constraints in the system. Before proceeding further, it is essential to note that we must specify any explicit (observer) time dependence. \textcolor{black}{This is because of the nonlinear nature of the Lagrangian~\eqref{eq:free-particle-lagrangian}, in the Hamiltonian formalism, we need to fix the reparametrization gauge (see, for example, Ref.~\cite{fulop1998reparametrization})\footnote{This gauge-fixing is not required if one uses the Lagrangian~\eqref{eq:free-particle-lagrangian2}.}} Let $t$ be the observer time and $x^{\alpha}(t)$ (with $\alpha$ ranging between $1$ and $4+m-1=3+m$) the coordinates parametrizing the motion in the embedding space $\mathcal{Q}$ at a time $t$.
We assume that after the reparametrization gauge is fixed, the Lagrangian $\mathcal{L}_{1}$ takes the symbolic form $\mathcal{L}_{0}(x^{\alpha}, \dot{x}^{\alpha}, t) + \Lambda^{i} \phi_{i}(x^{\alpha}, t)$ (where $\dot{x}^{\alpha} = dx^{\alpha}/dt$ is the velocity). 

\textcolor{black}{Henceforth, we shall use the Greek alphabets ($\alpha, \beta, \gamma \dots$) to refer to the original state-space variables  ($x^{\alpha}$) and the Latin alphabets ($i, j, \dots$) to index the constraints $\phi_{i}$ and the auxiliary state space variables $\Lambda_{i}$. We will refer to the momenta conjugate to $x^{\alpha}$ as $p_{\alpha}$ and to the momenta conjugate to $\Lambda^{i}$ as $\Pi_{i}$. We will use the symbol $\phi_{a, i}$ to refer to the $i^{\rm th}$ (with $i=1,\dots, m$) constraint of generation $a$ (with $a=1$ corresponding to the primary constraints, and $a=2$ to the secondary constraints and so on) and the symbol $\mu^{a, i}$ the corresponding Lagrange multipliers encountered in Dirac's procedure.}

The only non-invertible momenta in the enlarged state space are the momenta $\Pi_{i}$ conjugate to the auxiliary variables $\Lambda^{i}$. Therefore, the total Hamiltonian (in terms of the free-particle Hamiltonian $\mathcal{H}_{0}$) is defined in terms of the primary constraints $\phi_{1, i} \equiv \Pi_{i} \approx 0$ 
as
\begin{equation}\label{eq:total-hamiltonian}
  \mathcal{H}_{T} = \mathcal{H}_{0}(x^{\alpha}, p_{\alpha}, t) - \Lambda^{i}\phi_{i}(x^{\alpha}, t) + \mu^{1, i}\phi_{1, i}
\end{equation}

\subsection{Modified Poisson brackets}

The next step in Dirac's procedure is checking for the stability of the primary constraints $\phi_{1, i}$ against the total Hamiltonian $\mathcal{H}_{T}$. Doing so, straightforwardly yields $m$-secondary constraints $\phi_{2, i} \equiv \phi_{i} \approx 0$. Repeating the stability analysis for the secondary constraints $\phi_{2, i}$,
\begin{align*}
  \left\{\phi_{2, i},\mathcal{H}_{T}\right\} &= \left\{\phi_{i}(x^{\alpha}, t), \mathcal{H}_{T}\right\} = \frac{\partial}{\partial x^{\alpha}}\phi_{i}\frac{\partial}{\partial p_{\alpha}}\mathcal{H}_{T} \\
  &= \frac{\partial}{\partial x^{\alpha}}\phi_{i}\frac{\partial}{\partial p_{\alpha}}\mathcal{H}_{0}(x^{\beta}, p_{\beta})
\end{align*}
we find $m$ tertiary constraints $\phi_{3, i} \equiv \frac{\partial}{\partial x^{\alpha}}\phi_{i}(x^{\alpha})\frac{\partial}{\partial p_{\alpha}}\mathcal{H}_{0}(x^{\alpha}, p_{\alpha}) \approx 0$. Repeating Dirac's procedure (i.e., the stability analysis for the tertiary constraints $\phi_{3, i}$), 
\begin{align*}
  \left\{\phi_{3, i},\mathcal{H}_{T}\right\} &= \left\{\phi_{3, i}(x^\alpha, p_{\alpha}), \mathcal{H}_{0}\right\} - \Lambda^{j}\left\{\phi_{3, i}(x^{\alpha}, p_{\alpha}), \phi_{j}(x^{\alpha})\right\}\\
  &= \left\{\phi_{3, i}(x^\alpha, p_{\alpha}), \mathcal{H}_{0}\right\}  + \Lambda^{j}\frac{\partial}{\partial p_{\beta}}\phi_{3, i}(x^{\alpha}, p_{\alpha})\frac{\partial}{\partial x^{\beta}}\phi_{j}(x^{\alpha})\\
  &= \left\{\phi_{3, i}(x^\alpha, p_{\alpha}), \mathcal{H}_{0}\right\}  + \Lambda^{j}\frac{\partial}{\partial x^{\beta}}\phi_{i}(x^{\alpha})\frac{\partial}{\partial x^{\gamma}}\phi_{j}(x^{\alpha})\frac{\partial^{2}}{\partial p_{\beta}p_{\gamma}}\mathcal{H}_{0}
\end{align*}
we find the final generation of constraints $\phi_{4, i} \equiv \Lambda^{i} - \left(\left\{\phi_{3}, \phi_{2}\right\}\right)^{-1}_{i, j}\left\{\phi_{3, j}, \mathcal{H}_{0}\right\} \approx 0$ (because in general, $\vert \vert \frac{\partial}{\partial x^{\beta}}\phi_{i}(x^{\alpha})\frac{\partial}{\partial x^{\gamma}}\phi_{j}(x^{\alpha})\frac{\partial^{2}}{\partial p_{\beta}p_{\gamma}}\mathcal{H}_{0} \vert \vert \neq 0$). Here, the matrix $\left\{\phi_{3}, \phi_{2}\right\}$ is defined as $\left\{\phi_{3}, \phi_{2}\right\}_{i, j} = \left\{\phi_{3, i}, \phi_{2, j}\right\}$. Repeating Dirac's procedure,
\begin{align*}
  \left\{\phi_{4, i}, \mathcal{H}_{T}\right\} &= \left\{\Lambda_{i}, \mathcal{H}_{T}\right\} - \left\{\left(\left\{\phi_{3}, \phi_{2}\right\}\right)^{-1}_{i, j}\left\{\phi_{3, j}, \mathcal{H}_{0}\right\}, \mathcal{H}_{T}\right\} \\
  &= \left\{\Lambda^{i}, \mu^{1, j}\Pi_{j}\right\} - \left\{\left(\left\{\phi_{3}, \phi_{2}\right\}\right)^{-1}_{i, j}\left\{\phi_{3, j}, \mathcal{H}_{0}\right\}, \mathcal{H}_{0} + \Lambda^{k}\phi_{k}\right\} \\
  & = \mu^{1, i} - \left\{\left(\left\{\phi_{3}, \phi_{2}\right\}\right)^{-1}_{i, j}\left\{\phi_{3, j}, \mathcal{H}_{0}\right\}, \mathcal{H}_{0} + \Lambda^{k}\phi_{k}\right\}\\
  \because \phi_{4, i} &\equiv \Lambda^{i} \approx \left(\left\{\phi_{3}, \phi_{2}\right\}\right)^{-1}_{i, j}\left\{\phi_{3, j}, \mathcal{H}_{0}\right\}\\
  \implies \mu^{i} &\approx F^{i}(x^{\alpha}, p_{\alpha}) = \left\{\left(\left\{\phi_{3}, \phi_{2}\right\}\right)^{-1}_{i, j}\left\{\phi_{3, j}, \mathcal{H}_{0}\right\}, \mathcal{H}_{0} + (\left(\left\{\phi_{3}, \phi_{2}\right\}\right)^{-1}_{k, l}\left\{\phi_{3, l}, \mathcal{H}_{0}\right\})\phi_{k}\right\}
\end{align*}
we find that there are no further constraints.

We now calculate the Dirac brackets between the phase-space variables. Recall that the Dirac brackets between two functions $f, g$ on the phase space is defined as $\left\{f, g\right\}_{D}=\left\{f,g\right\}-\left\{f, \phi_{a, i}\right\}M^{-1}_{a, i; b, j}\left\{\phi_{b, j}, g\right\}$ where $M_{a, i; b, j} = \left\{\phi_{a, i}, \phi_{b, j}\right\}$. Therefore, we must calculate $M$ and its inverse $M^{-1}$ to evaluate the Dirac brackets.

In terms of matrices $A$, $B$, $C$ defined below, the matrix $M$ takes the following block structure (we specify the dimensions of the matrices as subscripts):
\begin{align*}
 M_{a, i; b, j} = \left\{\phi_{a, i}, \phi_{b, j}\right\} &= \begin{pmatrix}
    0_{m \times m} & 0_{m\times 2m} & -I_{m \times m} \\
    0_{2m \times m} & A_{2m \times 2m} & B_{2m \times m} \\
 I_{m \times m} & -B^{\rm T}_{m \times 2m} & C_{m \times m} \\
  \end{pmatrix} 
  \end{align*}
where $A \equiv \left\{\phi_{a, i},\phi_{b, j}\right\} \ni a, b = 2, 3$, $B \equiv \left\{\phi_{a, i},\phi_{4, j}\right\} \ni a = 2, 3$ and $C \equiv \left\{\phi_{4, i},\phi_{4, j}\right\}$.

The inverse of $M$ is equal to (from Schur's complement method for block matrices)
\begin{align*}
 M^{-1} &= \begin{pmatrix}
 C + B^{\rm T}A^{-1}B & B^{\rm T}A^{-1} & I \\
 A^{-1}B & A^{-1} & 0 \\
    -I & 0 & 0
  \end{pmatrix}
\end{align*}

Since, the phase space variables $\Lambda^{i}$ and $\Pi_{i}$ are constrained (by the primary and quaternary constraints $\phi_{1, i}$ and $\phi_{4, i}$) to be functions of $x^{\alpha}, p_{\alpha}$ (or are simply $0$), Dirac brackets of the form $\left\{\Lambda^{i}, f\right\}_{D}$ and $\left\{\Pi_{i}, f\right\}_{D}$ contain no additional information beyond the Dirac brackets $\left\{x^{\alpha}, f\right\}_{D}$ and $\left\{p_{\alpha}, f\right\}_{D}$. The Dirac brackets between functions of the phase space variables $x^{\alpha}, p_{\alpha}$ can be constructed by noting the following
\begin{align*}
  \left\{x^{\alpha}, \phi_{1, i}\right\} & = 0, \left\{p_{\alpha}, \phi_{1, i}\right\} = 0 \\
  \implies \left\{f(x^{\alpha}, p_{\beta}), \phi_{a, i}\right\}M^{-1}_{(a,i),(b,j)}\left\{\phi_{b, j}, g(x^{\gamma}, p_{\delta})\right\} &= \begin{pmatrix}
    0 & \left\{f, \phi_{(2, 3),i}\right\} & \left\{f, \phi_{4,i}\right\}
  \end{pmatrix} \times \\
  &\begin{pmatrix}
 C + B^{\rm T}A^{-1}B & B^{\rm T}A^{-1} & I \\
 A^{-1}B & A^{-1} & 0 \\
    -I & 0 & 0
  \end{pmatrix} \times \\
  & \begin{pmatrix}
    0 \\
    \left\{\phi_{(2, 3), j}, g\right\} \\
    \left\{\phi_{4, j}, g\right\}
  \end{pmatrix} \\
  &= \left\{f, \phi_{(2, 3), i}\right\}A^{-1}\left\{\phi_{(2, 3), j}, g\right\}
\end{align*}
The Dirac brackets between the phase space variables $x^{\alpha}, p_{\alpha}$ is therefore equal to:
\begin{equation}\label{eq:classical-dirac-brackets}
  \begin{split}
    \left\{x^{\alpha}, x^{\beta}\right\}_{D} &= -\left\{x^{\alpha}, \phi_{a, i}\right\}A^{-1}_{(a,i),(b,j)}\left\{\phi_{b, j}, x^{\beta}\right\}\\
    \left\{x^{\alpha}, p_{\beta}\right\}_{D} &= \delta^{\alpha}_{\beta}-\left\{x^{\alpha}, \phi_{a, i}\right\}A^{-1}_{(a,i),(b,j)}\left\{\phi_{b, j}, p_{\beta}\right\}\\
    \left\{p_{\alpha}, p_{\beta}\right\} &= -\left\{p_{\alpha}, \phi_{a, i}\right\}A^{-1}_{(a,i),(b,j)}\left\{\phi_{b, j}, p_{\beta}\right\}\\
 A &= \left\{\phi_{a, i}, \phi_{b, j}\right\} \ni a, b = 2, 3
  \end{split}
\end{equation}
{We see that the matrices $\left\{x^{\alpha}, \phi_{a, i}\right\}$, $\left\{p_{\alpha}, \phi_{a, i}\right\}$ and $A = \left\{\phi_{a, i}, \phi_{b, j}\right\}$ involving the secondary and tertiary constraints $\phi_{(2, 3), i}$ completely determine the Dirac bracket structure for the constrained motion in the embedding space $\mathcal{Q}$. Since the matrices $\left\{x^{\alpha}, \phi_{a, i}\right\}$ and $\left\{p_{\alpha}, \phi_{a, i}\right\}$ play a vital role in the modifications to the standard position-momentum algebra, we shall refer to them as \textit{building blocks}}.

Note that the other (auxiliary) phase space variables $\Lambda^{i}, \Pi_{i}$ are constrained to be functions of $x^{\alpha}, p_{\alpha}$ or constrained to be $0$ and hence do not contribute to the dynamics.

With the Dirac brackets determined for the classical constrained motion, i.e., for the motion of a point particle in a $4$-D curved space-time $\mathcal{P}$, we now follow Dirac and propose that the quantum mechanical dynamics of a point particle in a curved space-time $\mathcal{P}$ is described by the Hamiltonian
\begin{equation}\label{eq:final-hamiltonian}
  \hat{H}(\hat{x}^{\alpha}, \hat{p}_{\alpha}, t) = \hat{H}_{0}[\hat{x}^{\alpha}, \hat{p}_{\alpha}, t] + c_{a, i}\hat{\mu}^{a, i}
\end{equation}
Here, $\hat{H}_{0}$ is the Hamiltonian of the particle in the flat embedding space $\mathcal{Q}$, \textcolor{black}{$c_{a, i}$ are arbitrary $c$-numbers} (due to the promotion of the constraints $\phi_{a, i} \approx 0$ to $\hat{\phi}_{a, i}[\hat{x}^{\alpha}, \hat{p}_{\alpha}, \hat{\Lambda}^{i}, \hat{\Pi}_{i}] = c_{a, i}\mathbb{I}$) that contribute to dynamics at $\mathcal{O}(\hbar^{2})$ and $\hat{\mu}^{a, i}$ are operators that cannot be determined within the framework of Dirac's procedure. 

The key here is that the operators $\hat{x}^{\alpha}$ and $\hat{p}_{\alpha}$ obey a modified position-momentum algebra (i.e., the operator version of Eq.~\ref{eq:classical-dirac-brackets})
\begin{equation}\label{eq:dirac-brackets-final}
  \begin{split}
    \left[\hat{x}^{\alpha}, \hat{x}^{\beta}\right] &= -\imath \hbar \left(\left\{x^{\alpha}, \phi_{a, i}\right\}A^{-1}_{(a,i),(b,j)}\left\{\phi_{b, j}, x^{\beta}\right\}\right)_{x^{\gamma} \rightarrow \hat{x}^{\gamma}, p_{\delta} \rightarrow \hat{p}_{\delta}}\\
    \left[\hat{x}^{\alpha}, \hat{p}_{\beta}\right] &= \imath \hbar \delta^{\alpha}_{\beta}\hat{I}-\imath \hbar \left(\left\{x^{\alpha}, \phi_{a, i}\right\}A^{-1}_{(a,i),(b,j)}\left\{\phi_{b, j}, p_{\beta}\right\}\right)_{x^{\gamma} \rightarrow \hat{x}^{\gamma}, p_{\delta} \rightarrow \hat{p}_{\delta}} \\
    \left[\hat{p}_{\alpha}, \hat{p}_{\beta}\right] &= -\imath \hbar \left(\left\{p_{\alpha}, \phi_{a, i}\right\}A^{-1}_{(a,i),(b,j)}\left\{\phi_{b, j}, p_{\beta}\right\}\right)_{x^{\gamma} \rightarrow \hat{x}^{\gamma}, p_{\delta} \rightarrow \hat{p}_{\delta}}\\
 a & = 2, 3
  \end{split}
\end{equation}
and satisfy the constraints
\begin{equation}\label{eq:constraints-final}
  \begin{split}
    \hat{\phi}_{2, i} &= \phi_{i}[\hat{x}^{\alpha}, t] = c_{2}\hat{I} \\
    \hat{\phi}_{3, i} &= \frac{\partial}{\partial y^{\alpha}}\phi_{i}(y^{\alpha})\vert_{y^{\alpha} \rightarrow \hat{x}^{\alpha}}\frac{\partial}{\partial p_{\alpha}}H_{0}(\hat{x}^{\alpha}, p_{\alpha})\vert_{p_{\alpha} \rightarrow \hat{p}_{\alpha}} = c_{3}\hat{I}
  \end{split}
\end{equation}
Here, the RHS of equations Eq.~\eqref{eq:dirac-brackets-final} and Eq.~\eqref{eq:constraints-final} are evaluated classically and then promoted into operators. The other two constraints fix $\hat{\Pi}_{i}$ and $\hat{\Lambda}^{i}$ as functionals of $\hat{x}^{\alpha}$ and $\hat{p}_{\alpha}$, and are therefore ignored.

Thus, the dynamics of a relativistic particle in arbitrary 4-D space-time as described by the Lagrangian \eqref{eq:constrained-lagrangian} can be described by embedding the 4-D curved geometry in a higher-dimensional flat embedding space. As mentioned in the Introduction, this procedure does not require the notion of asymptotically flat space-time or the definition of a plane wave in the asymptotic limit.

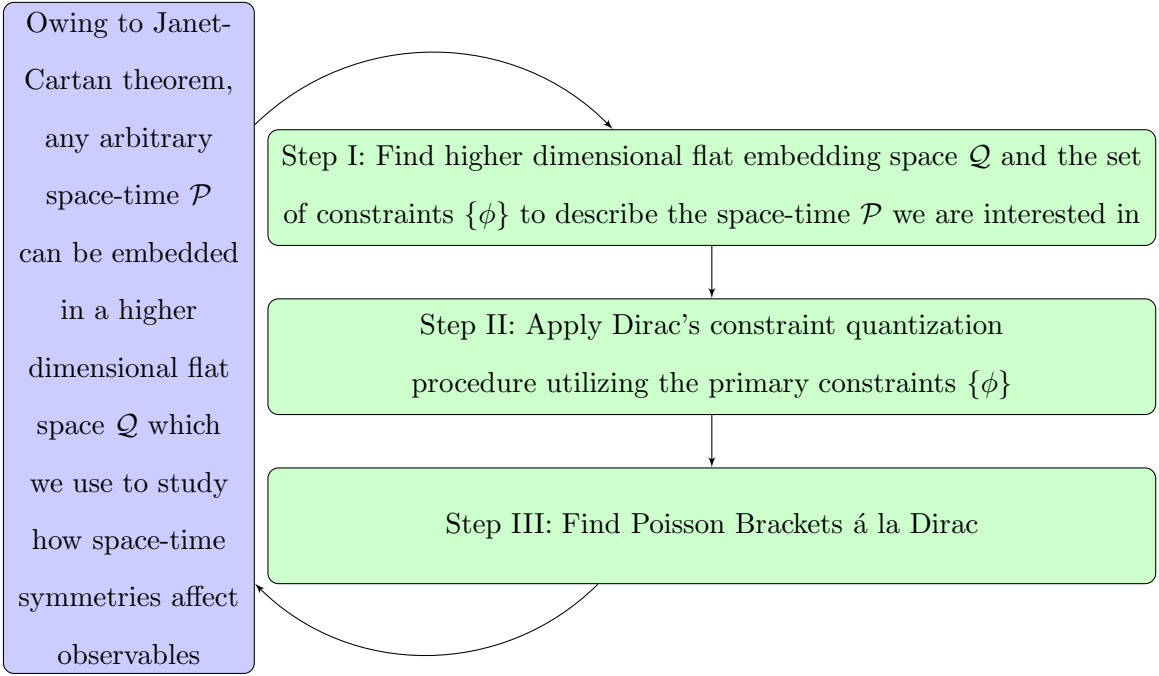
\begin{figure*}
  \begin{tikzpicture}[node distance = 2.25cm, auto]
    \node [blockkk] (B11) {Owing to Janet-Cartan theorem, any arbitrary space-time $\mathcal{P}$ can be embedded in a higher dimensional flat space $\mathcal{Q}$ which we use to study how space-time symmetries affect observables};
    \node [block, left of=B11, xshift=10.0cm, yshift=2cm] (B12) {Step I: Find higher dimensional flat embedding space $\mathcal{Q}$ and the set of constraints $\left\{\phi\right\}$ to describe the space-time $\mathcal{P}$ we are interested in};
    \node [block, below of=B12] (B13) {Step II: Apply Dirac’s constraint quantization procedure};
    \node [block, below of=B13] (B14) {Step III: Find Poisson Brackets \'{a} la Dirac};
    \path [line] (B11) edge[bend left=45] node [right]{}(B12);
    \path [line] (B12) -- (B13);
    \path [line] (B13) -- (B14);
  \end{tikzpicture}
  \caption{\small A systematic procedure for deriving modified position-momentum algebras reflecting the description of quantum mechanics in curved geometries.} \label{fig:Flowchart}
\end{figure*}

In summary, we use Dirac's constraint quantization procedure to relate the symmetries of the 4-dimensional space-time with the modification of the position-momentum algebra. In the next section, we apply our procedure for a few space-times whose embeddings are known~\cite{10.1007/3-540-46671-1_5}.
To wrap up the discussion for arbitrary 4-D space-time, we list below the steps involved:
\begin{STEP}
  \item We identify the optimal (flat) embedding space $\mathcal{Q}$ for the 4-dimensional curved space-time $\mathcal{P}$ of interest and the corresponding set of constraints $\left\{\phi_{a}\right\}$.
  \item We enlarge our state space to include the Lagrange multiplier variables $\left\{\mu_{a}\right\}$ that we use to impose the holonomic constraints $\left\{\phi_{a}\right\}$. We then follow Dirac's constraint procedure to derive the Hamiltonian dynamics. 
  \item Using Dirac Brackets, we incorporate the effects of the constrained dynamics into a new generalized Poisson Brackets and construct the algebra for canonical variables in the total Hamiltonian $\mathcal{H}_{T}$. Finally, we construct the quantum mechanical position-momentum algebra.
\end{STEP}

Figure \eqref{fig:Flowchart} provides a bird's eye view of the steps followed.

\section{Application to specific space-times}
\label{sec:applications}

In this section, we apply the procedure discussed in the previous section for a few space-times whose embeddings are known~\cite{10.1007/3-540-46671-1_5} and show that the corrections in the position-momentum algebra are proportional to the curvature invariants of the curved space-times.

\subsection{Anti-de Sitter space-time}
\label{sec:AdS}

The $(1,3)$ anti-de Sitter (AdS) space-time $\mathcal{P}$ can be embedded in a $(2,3)$ flat space $\mathcal{Q}$ as follows:
\begin{equation}\begin{split}ds^{2} & = -\left(dX^{0}\right)^{2}-\left(dX^{1}\right)^{2}+\sum_{A=2}^{4}\left(dX^{A}\right)^{2} \\
    \phi_{1}& \equiv 
\left(X^{0}\right)^{2}+\left(X^{1}\right)^{2}-\sum_{A=2}^{4}\left(X^{A}\right)^{2}-\alpha^{2} \approx 0 \, , \\ \end{split}\end{equation}
where \textcolor{black}{$\alpha \in \mathbb{R}^{+}$} is related to the negative cosmological constant. In terms of the observer time $t$, we take $\mathcal{Q}$ to be parametrized as follows
\begin{equation}\label{AdS-time-dependence}\begin{split}
  X^{0} & = x^{0}\cos t \\
  X^{1} & = x^{0}\sin t \\
  X^{\alpha+1} & = x^{\alpha} , \alpha \in \left\{1,\dots,3\right\}\\
\end{split}\end{equation}
For notational ease, we introduce a diagonal matrix $n_{\alpha \beta} = \mathrm{Diag}(1, -1, -1, -1)$ such that, the Hamiltonian $\mathcal{H}_{0}$ and the constraint $\phi_{1}$ can be written as
\begin{equation}
  \begin{split}
    \mathcal{H}_{0} &= x^{0}\sqrt{m_{0}^{2}-n^{\alpha \beta}p_{\alpha}p_{\beta}} \\
    \phi_{1}(x^{\alpha}) &= n_{\alpha \beta}x^{\alpha}x^{\beta}-\alpha^2 \approx 0
  \end{split}
\end{equation}

Following the procedure outlined in Sec.~\eqref{sec:Procedure}, we first construct the set of secondary and tertiary constraints $\phi_{(2, 3), i}$ (see Eq.~\eqref{eq:constraints-final}) as
\begin{equation}
  \begin{split}
    \phi_{2, 1} &= n_{\alpha \beta}x^{\alpha}x^{\beta}-\alpha^2 \approx 0 \\
    \phi_{3, 1} &= x^{\alpha}p_{\alpha} \approx 0
  \end{split}
\end{equation}

We then calculate the Poisson brackets matrix $A$ between the secondary and tertiary constraints,
\begin{equation}
  A = \begin{pmatrix}
    \left\{\phi_{2, 1}, \phi_{2, 1}\right\} & \left\{\phi_{2, 1}, \phi_{3, 1}\right\}\\
    \left\{\phi_{3, 1}, \phi_{2, 1}\right\} & \left\{\phi_{3, 1}, \phi_{3, 1}\right\}\\
  \end{pmatrix} = \begin{pmatrix}
    0 & 2\alpha^{2}\\
    -2\alpha^{2} & 0 \\
  \end{pmatrix}
\end{equation}
its inverse
\begin{equation}\label{eq:inv-A-AdS}
  A^{-1}  = \begin{pmatrix}
    0 & -\frac{1}{2\alpha^{2}}\\
    \frac{1}{2\alpha^{2}} & 0 \\
  \end{pmatrix} \, .
\end{equation}
The building blocks $\left\{x^{\alpha}, \phi_{(2, 3), i}\right\}$ and $\left\{p_{\alpha}, \phi_{(2, 3), i}\right\}$ satisfy the following relations:
\begin{equation}\label{eq:building-blocks-AdS}
  \begin{split}
    \left\{x^{\alpha}, \phi_{2, 1}\right\} &=0, \left\{x^{\alpha}, \phi_{3, 1}\right\} =x^{\alpha}\\
    \left\{p_{\alpha}, \phi_{2, 1}\right\} &=-2x_{\alpha}, \left\{p_{\alpha}, \phi_{3, 1}\right\} =-p_{\alpha}\\
  \end{split}
\end{equation}

Substituting Eqs.~\eqref{eq:inv-A-AdS} and \eqref{eq:building-blocks-AdS} in Eq.~\ref{eq:dirac-brackets-final}, we construct the Dirac brackets between $x^{\alpha}$ and $p_{\beta}$:
\begin{align*}
    \left\{x^{\alpha}, x^{\beta}\right\}_{D} &= 0\\
    \left\{x^{\alpha}, p^{\beta}\right\}_{D} &= n^{\alpha \beta} - \frac{x^{\alpha}x^{\beta}}{\alpha^{2}} \\
    \left\{p^{\alpha}, p^{\beta}\right\}_{D} &= -\frac{1}{\alpha^{2}}\left(x^{\alpha}p^{\beta}-x^{\beta}p^{\alpha}\right)
\end{align*}
Hence, the quantum dynamics of a point particle in AdS space-time, can be described in terms of a modified position-momentum algebra
\begin{equation}\label{eq:AdS-Dirac-Op}
  \begin{split}
    \left[\hat{x}^{\alpha}, \hat{x}^{\beta}\right] &= 0\\
    \left[\hat{x}^{\alpha}, \hat{p}^{\beta}\right] &= \imath \hbar n^{\alpha \beta}\mathbb{I} -\imath \hbar \frac{1}{\alpha^{2}}\hat{x}^{\alpha}\hat{x}^{\beta}\\
    \left[\hat{p}^{\alpha}, \hat{p}^{\beta}\right] &= -\imath \hbar \frac{1}{\alpha^{2}}\left(\hat{x}^{\alpha}\hat{p}^{\beta}-\hat{x}^{\beta}\hat{p}^{\alpha}\right)\\
  \end{split}
\end{equation}
with the Hamiltonian (up to leading order i.e. we set the complex numbers $c_{i}$ in Eq.~\ref{eq:final-hamiltonian} to $0$)
\begin{equation}
  \hat{H}_{\mathrm{T}} = \hat{x}^{0}\sqrt{m_{0}^{2}-\hat{p}^{\alpha}\hat{p}_{\alpha}}
\end{equation}

This is the first key result of this work, regarding which we would like to stress the following: First, in the flat space-time limit (i.e., $\alpha \rightarrow \infty$), we recover the standard Heisenberg algebra:
\begin{equation}\label{AdS-Flat-Space}
\begin{split}
    \left[\hat{x}^{\alpha},\hat{x}^{\beta}\right] & = \left[p^{\alpha},p^{\beta}\right] = 0\\
    \left[\hat{x}^{\alpha},\hat{p}^{\beta}\right] & = \imath \hbar n^{\alpha \beta}\mathbb{I}\\
\end{split}
\end{equation}
However, note that the embedding strictly does not exist when ${1}/{\alpha} \to 0$. 
Second, the corrections to the standard commutation relations (Eq.~\ref{eq:AdS-Dirac-Op}) are proportional to curvature invariants. In this case, Ricci scalar $(R)$ is $-12/\alpha^2$. As we show for other space-times, the corrections to the standard commutation relations are always related to curvature invariants. 
Third, the corrections to the standard commutation relations are significant when $\langle \hat{x}^{\alpha} \hat{x}^{\beta} \rangle$ is comparable to $\alpha^2$. This shows that the corrections are significant in the infrared limit. 
Fourth, using Eq.~\eqref{eq:AdS-Dirac-Op}, we can find all observables that pertain to a particle moving in anti-de Sitter space-time by expressing them in terms of the embedding coordinates $x^{\alpha}$ and $p^{\alpha}$.
Finally, the results here match those obtained in the literature~\cite{mignemi2010extended,mallett1973position}. This validates our procedure and provides a new way of looking at the curvature effects on the position-momentum algebra.

\subsection{de Sitter space-time}

The static patch of the de Sitter space-time is:
\begin{equation}\label{dS-metric}
\begin{split}
ds^{2} = \left(1- \frac{r^{2}}{\alpha^{2}} \right) d t^{2} - \frac{1}{(1-r^{2} / \alpha^{2})} d r^{2} - r^{2} d\Omega^2 \,  ,
          \end{split}
        \end{equation}
where $\alpha \in \mathbb{R}^{+}$ is related to the positive cosmological constant and $d\Omega^2$ is the {line-element} on the unit sphere $S^2$. The above line-element can be embedded in a $(1,4)$ flat space as follows~\cite{rosen1965embedding}:
\begin{equation}\label{dS-embedding}
\begin{split}
X^{0} = x^{0}\sinh \frac{t}{\alpha};~~~ 
X^{1}  = x^{0}\cosh \frac{t}{\alpha}; &
~~~X^{\alpha+1}  = x^{\alpha} \ni \alpha~~ \left\{1,2,3\right\}\\
\phi_{1} \equiv \delta_{\alpha \beta}x^{\alpha}x^{\beta} -\alpha^{2} = 0 & \\
          \end{split}
        \end{equation}
For notational ease, we have introduced the identity matrix $\delta_{\alpha \beta} \equiv \mathrm{Diag}(1, 1, 1, 1)$ to raise and lower indices. Repeating the same steps as in the case of AdS space-time, we obtain the following position-momentum algebra:
        \begin{equation}\label{dS-Dirac-Op}
          \begin{split}
            \left[\hat{x}^{\alpha},\hat{x}^{\beta}\right] & = 0\\
           \left[\hat{x}^{\alpha},\hat{p}^{\beta}\right] & = \imath \hbar \left(\delta^{\alpha \beta}\id-\frac{1}{\alpha^{2}}\hat{x}^{\alpha}\hat{x}^{\beta}\right) \\
 \left[\hat{p}^{\alpha},\hat{p}^{\beta}\right] & = -\frac{\imath \hbar}{\alpha^{2}} \left(\hat{x}^{\alpha}\hat{p}^{\beta}-\hat{x}^{\beta}\hat{p}^{\alpha}\right) \\
        \end{split}
    \end{equation}

We want to make the following remarks regarding the above result: First, like in the case of AdS space-time, we recover the standard Heisenberg algebra in the flat space-time limit ($\alpha \to \infty$). 
Second, the corrections to the standard commutation relations \eqref{dS-Dirac-Op} are proportional to Ricci scalar $R = 12/\alpha^2$. 
Third, the position-momentum algebra for AdS \eqref{eq:AdS-Dirac-Op} and dS \eqref{dS-Dirac-Op} may appear identical; however, in terms of the scalar invariant, the correction terms have an overall sign difference. Thus, our results show that the curvature of the space-time is directly related to the modified position-momentum algebra.

\subsection{Schwarzschild space-time}

The \textcolor{black}{Schwarzschild line-element} is given by:
\begin{equation}\label{Schwarzschild-metric}
\begin{split}
ds^{2}= h(r) d t^{2} - \frac{dr^2}{h(r)}- r^{2} d\Omega^2;~~ h(r) = 1- \frac{r_{_{S}}}{ r}
\end{split}
\end{equation}
where $r_S = 2 G M$ and $M$ is the mass. The above line-element can be embedded in a $(2,4)$ flat space as follows~\cite{rosen1965embedding}:
\begin{alignat}{2}\label{Schwarzschild-embedding}
\nonumber & X^{0} = r_{_{S}} \sqrt{h(r)} 
\cos \frac{t}{r_{_{S}}} = x^{0}\cos  \frac{t}{r_{_{S}}} \;,X^{3} = r \sin \theta \cos \phi = x^{2}\\ 
\nonumber & X^{1} = r_{_{S}} \sqrt{h(r)} \sin \frac{t}{r_{_{S}}} = x^{0} \sin  \frac{t}{r_{_{S}}} \;, X^{4} = r \sin \theta \sin \phi = x^{3}\\
\nonumber & X^{2} = \int dr \sqrt{
\frac{4+h^{\prime}(r)^{2}}{4h(r)}-1} = f(x^{0}) \equiv x^{1}
\,, X^{5} = r \cos \theta = x^{4}\\          
& \phi_{1} \equiv \sum_{\alpha=2}^{4}(x^{\alpha})^{2} - g\left(x^{0}\right)^{2} \approx 0\;,\; \phi_{2}\equiv x^{1}-f(x^{0}) \approx 0
\end{alignat}
\textcolor{black}{Here, $g$ and $f$ are functions such that, $r = g(r_{s}\sqrt{h(r)})$ and $\int dr \sqrt{
  \frac{4+h^{\prime}(r)^{2}}{4h(r)}-1} = f(r_{s}\sqrt{h(r)})$. In practice, we expect these functions to be determined numerically.} Repeating the same steps as in AdS and dS space-times, we obtain the following position-momentum algebra:  
\begin{equation*}\label{Schwarzschild-dirac-op}
          \begin{split}
            & \left[\hat{x}^{\alpha},\hat{x}^{\beta}\right] = 0 \\
            & \left[\hat{x}^{0}, \hat{p}_{0}\right] = -\frac{\imath \hbar}{4}\left(1-\frac{\left(\hat{x}^{0}\right)^{2}}{r_{_{S}}^{2}}\right)^{4},\left[\hat{x}^{0}, \hat{p}_{1}\right] = -f^{\prime}(\hat{x}^{0})\left[\hat{x}^{0},\hat{p}_{0}\right]\\
            & \left[\hat{x}^{1}, \hat{p}_{0}\right] = f^{\prime}(\hat{x}^{0})\left[\hat{x}^{0},\hat{p}_{0}\right],\left[\hat{x}^{1}, \hat{p}_{1}\right] = -f^{\prime}(\hat{x}^{0})^{2}\left[\hat{x}^{0},\hat{p}_{0}\right]\\
            & \left[\hat{x}^{0}, \hat{p}_{\alpha}\right] = \frac{\imath \hbar r_{_{S}}}{2\hat{r}^{3}}\hat{x}^{0}\hat{x}^{\alpha},\left[\hat{x}^{1}, \hat{p}_{\alpha}\right] = f^{\prime}(\hat{x}^{0})\left[\hat{x}^{0}, \hat{p}_{\alpha}\right] \\
            & \left[\hat{x}^{\alpha}, \hat{p}_{0}\right] = -\frac{\imath \hbar r_{_{S}}}{2\hat{r}^{3}}\hat{x}^{0}\hat{x}^{\alpha},\left[\hat{x}^{\alpha}, \hat{p}_{1}\right] = -f^{\prime}(\hat{x}^{0})\left[\hat{x}^{\alpha}, \hat{p}_{0}\right] 
    \end{split}
        \end{equation*}
  \begin{equation}\label{Schwarzschild-dirac-op2}
          \begin{split}          
            &\left[\hat{x}^{\alpha},\hat{p}_{\beta}\right] = \imath \hbar \left(\delta^{\alpha}_{\beta}\id - \frac{r_{_{S}}}{\hat{r}^{3}}\hat{x}^{\alpha}\hat{x}_{\beta}\right)\\
            &\left[\hat{p}_{0}, \hat{p}_{1}\right] = f^{\prime \prime}(\hat{x}^{0})\left[\hat{x}^{0},\hat{p}_{0}\right]\hat{p}_{0}\\
            & \left[\hat{p}_{0}, \hat{p}_{\alpha}\right] = -\imath \hbar\frac{r_{_{S}}^{4}}{4\hat{r}^{6}}\left(\frac{\partial}{\partial x^{0}}\left(gg^{\prime}\right)\hat{x}_{\alpha}\hat{p}_{0} + gg^{\prime}\hat{p}_{\alpha}\right)\\
            & \left[\hat{p}_{1}, \hat{p}_{\alpha}\right] = -f^{\prime}(\hat{x}^{0})\left[\hat{p}_{0},\hat{p}_{\alpha}\right]-\imath \hbar \frac{r_{_{S}}^{4}gg^{\prime}f^{\prime \prime}}{4\hat{r}^{6}}\hat{x}^{\alpha}\hat{p}_{0}\\
            & \left[\hat{p}_{\alpha}, \hat{p}_{\beta}\right] = -\frac{\imath \hbar r_{_{S}}}{\hat{r}^{3}}\left(\hat{x}_{\alpha}\hat{p}_{\beta}-\hat{x}_{\alpha}\hat{p}_{\beta}\right) \;,\; \alpha,\beta\in \left\{2,3,4\right\} 
          \end{split}
        \end{equation}
This is the second key result of this work, regarding which we would like to stress the following: First, in the flat space-time limit (i.e., $r_{_S} \to 0$), we have two independent sub-algebras. The corrections to one of the sub-algebras formed by $\hat{x}^{0},\hat{x}^{1},\hat{p}_{0},\hat{p}_{1}$ do not vanish in this limit. We attribute this to the fact that $x^0$ and $x^1$ always have a finite range in this limit. 
However, the second sub-algebra formed by $\hat{x}^2, \hat{x}^3, \hat{x}^4, \hat{p}_2, \hat{p}_{3}, \hat{p}_{4}$, reduces to the standard Heisenberg algebra. Unlike $x^{0},x^{1}$ the coordinates $x^{2},x^{3},x^{4}$ have infinite ranges for all values of $r_{_S}$. This establishes that the structure of the sub-algebras formed is dependent on the ranges of the coordinates. 
Second, the corrections to this sub-algebra formed by $\hat{x}^2, \hat{x}^3, \hat{x}^4, \hat{p}_2, \hat{p}_{3}, \hat{p}_{4}$ are proportional to the Kretschmann scalar:
\begin{equation}
K(r) = R_{\mu\nu\alpha\beta} R^{\mu\nu\alpha\beta} = 
\frac{12 r_{_S}^2}{r^6}
\end{equation}
At the horizon radius of a Solar mass black hole, the square root of the Kretschmann scalar is $10^{-6}~m^{-2}$. However, at the horizon radius of a primordial black hole (PBH) of mass, $10^{-3} M_{\odot}$,  the square root of Kretschmann scalar is $3~m^{-2}$~\cite{2022-Johnson.Shanki-GRG}. Thus, our analysis implies that the modified position-momentum algebra will be relevant and provide non-trivial corrections to PBH. 
Third, although we have explicitly derived the modified position-momentum algebra for \textcolor{black}{Schwarzschild space-time}, the analysis can be extended to any generic spherically symmetric space-time of the form: 
\begin{equation}\label{Spher-metric}
\begin{split}
ds^{2}= h_1(r) d t^{2} - \frac{dr^2}{h_2(r)}- r^{2} d\Omega^2 \, ,
\end{split}
\end{equation}
where $h_1(r) \neq h_2(r)$. 

\subsection{Geon space-time}

The static and cylindrically symmetric Geon is described by~\cite{rosen1965embedding}:
\begin{equation}\label{geon-metric}
\!\!\! ds^{2} = A^{2}(r) \left[d t^{2}-d r^{2}-d z^{2}\right] - \frac{r^{2} d \phi^{2}}{A^{2}(r)};~~A(r) = 1 + \frac{r^2}{a^2}
\end{equation}
Here, $a$ is the range radius of the flux structure and is related to the magnetic field. The above line-element can be embedded in a $(3,4)$ flat space as follows~\cite{rosen1965embedding}:
\begin{equation*}\label{geon-embedding}
\begin{split}
&X^{0}= A(r) \cos t = x^{0}\cos t \; , \; X^{1} = A(r) \sin t =  x^{0}\sin t,\\
&X^{2} = \frac{r}{A(r)}\cosh \phi = x^{1}\; , \; 
X^{3} = \frac{r}{A(r)}\sinh \phi = x^{2}\\
&X^{4} = A(r) \cos z = x^{3}\; , \; 
X^{5} = A(r) \sin z = x^{4}\\
     \end{split}
\end{equation*}
\begin{equation}\label{geon-embedding2}
\begin{split}
&X^{6} =\int dr \sqrt{A^2(r) + \frac{\left(1-r^{2} / a^{2}\right)^{2}}{A^4(r)}} = f(x^{0}) = x^{5}\\
& \phi_{1} \equiv \left(x^{1}\right)^{2}-\left(x^{2}\right)^{2} = g(x^{0})^{2}\;,\;\phi_{2} \equiv \left(x^{3}\right)^{2}+\left(x^{4}\right)^{2} = \left(x^{0}\right)^{2}\\
&\phi_{3} \equiv x^{5}-f(x^{0}) \approx 0
     \end{split}
\end{equation}
\textcolor{black}{Here, $f$ and $g$ are functions such that, $f(A(r)) = \int dr \sqrt{A^{2}(r) + (1-r^2/a^{2})^{2}/A(r)^4}$ and $g(A(r)) = r/A(r)$. Again, we expect that, in practice, the functions $f$ and $g$ are determined numerically.}

Repeating the steps as in (anti-)de Sitter and Schwarzschild space-times, we obtain the \emph{building blocks} and the position-momentum algebra for the Geon space-time. Here, we present the building blocks, not the position-momentum algebra, since they are more illuminating:
        \begin{equation}\label{geon-pos-mom-matrices}
          \begin{split}
            \left\{x^{\alpha}, \phi_{2, i}\right\} & = \begin{pmatrix}
              0 & 0 & 0\\
              0 & 0 & 0\\
              0 & 0 & 0\\
              0 & 0 & 0\\
              0 & 0 & 0\\
              0 & 0 & 0\\
            \end{pmatrix}~~~\left\{x^{\alpha}, \phi_{3, i}\right\}= \begin{pmatrix}
              -gg^{\prime} & x^{0} & f^{\prime}\\
 x^{1} & 0 & 0 \\
 x^{2} & 0 & 0 \\
              0 & x^{3} & 0 \\
              0 & x^{4} & 0 \\
              0 & 0 & 0\\
            \end{pmatrix}\\
            \left\{p_{\alpha}, \phi_{2, i}\right\} & = \begin{pmatrix}
              2gg^{\prime} & -2x^{1} & 2x^{2}\\
              2x^{0} & 0 &0\\
 f^{\prime} & 0 & 0\\
 p_{0}\left(gg^{\prime}\right)^{\prime} & -p_{1} & -p_{2}\\
              -p_{0} & 0 & 0\\
              -p_{0}f^{\prime \prime} & 0 & 0\\
            \end{pmatrix}~~~\left\{p_{\alpha}, \phi_{3, i}\right\} = \begin{pmatrix}
              0 & 0 & 0 \\
              -2x^{3} & -2x^{4} & 0 \\
              0 & 0 & -1 \\
              0 & 0 & 0 \\
              -p_{3} & -p_{4} & 0 \\
              0 & 0& 0 \\
            \end{pmatrix}
          \end{split}
        \end{equation}
We want to make the following remarks regarding the above result: 
First, it's crucial to note that in the flat space-time limit ($a \to \infty$), the coordinates with finite ranges  [$x^{0},x^{1},x^{2},x^{5},p_{0},p_{1},p_{2},p_{5}$] form sub-algebras that remain distinct from the standard Heisenberg algebra. Conversely, the coordinates with infinite ranges  [$x^{3},x^{4},p_{3},p_{4}$] form sub-algebras that reduce to the standard Heisenberg algebra in the flat space-time limit.
Second, we see that the symmetry of space-time plays a crucial role in the structure of the position-momentum algebras and sub-algebras. We want to make note of the following:
\begin{enumerate}[label=\alph*)]
    \item Sub-groups with spherical symmetry $SO(n)$, have the following structure,
    \begin{equation}\label{spher-symmetry-sub-group}
        \begin{split}
            \left[\hat{x}^{\alpha},\hat{x}^{\beta}\right] & = 0\\
            \left[\hat{x}^{\alpha},\hat{p}_{\beta}\right] & = \imath \hbar \left(\delta^{\alpha}_{\beta}\id -\frac{1}{r(\hat{x})^{2}}x^{\alpha}x^{\beta}\right)\\
            \left[\hat{p}_{\alpha},\hat{p}_{\beta}\right] & = -\frac{\imath \hbar}{r^{2}(\hat{x})}\left(\hat{x}^{\alpha}\hat{p}_{\beta}-\hat{x}^{\beta}\hat{p}_{\alpha}\right)\\
        \end{split}
    \end{equation}
Here, $r$ is the radial coordinate. Sub-groups with symmetries such as $SO(d,m)$, exhibit a structure similar to \eqref{spher-symmetry-sub-group}, with the time-like coordinates replacing $\hat{x}\rightarrow -\hat{x}$ and $\hat{p} \rightarrow -\hat{p}$. The radial coordinate ($r$) does not explicitly depend on $x^{\alpha}$s, and therefore, we expect the above result to hold for any sub-group of symmetry $SO(d,m)$. For instance, we see this in AdS  \eqref{eq:AdS-Dirac-Op}, dS \eqref{dS-Dirac-Op} and Schwarzschild \eqref{Schwarzschild-dirac-op} space-times. 
    
 \item Another key observation is that anisotropy in $4$-dimensional space-time is directly mirrored in the structure of the position-momentum algebra. For instance, in the cylindrically symmetric Geon, the presence of two different subgroups with symmetry of the form $SO(d,m)$ clearly reflects this anisotropy. 

\item The inhomogeneity of $4$-dimensional space-time is also reflected in the modified position-momentum commutation relations. We can understand this by focusing on the differences between the results obtained for the homogeneous space-times, like AdS~\eqref{eq:AdS-Dirac-Op} and  dS~\eqref{dS-Dirac-Op} and non-homogeneous space-times, like Schwarzschild \eqref{Schwarzschild-dirac-op} and Geon \eqref{geon-pos-mom-matrices}.
\end{enumerate}

\section{Conclusions and Discussions}
\label{sec:conc}

This work explicitly shows that the infrared effects in quantum physics arise generically in curved space-time. More specifically, we have demonstrated that the modifications to the position-momentum algebra are proportional to the curvature invariants (like the Ricci scalar and the Kretschmann scalar). Thus, our results show that infrared effects in quantum systems do not require any extra assumptions. This has to be contrasted to UV modifications in the quantum system, where 
we must make assumptions about the Planck scale physics. Although we have applied the procedure to a few space-times, the method generally applies to any space-time whose embedding is known. Thus, the procedure is robust and does not require any specific symmetry. 

Understanding field theories in AdS and dS space-times 
has necessitated the introduction of infrared cutoff~\cite{Bolen:2004sq,Park:2007az}. Given the conjugate nature of position and momentum in Quantum Mechanics, it is expected that quantum gravity should give rise to non-commutative momentum space --- a minimum momentum scale and corrections to the Heisenberg Uncertainty Principle \eqref{eup}. However, 
none of the earlier approaches~\cite{kempf1996noncommutative,Bolen:2004sq,Park:2007az,Bambi:2007ty,costa2016extended,wagner2022relativistic, petruzziello2021gravitationally} could derive the extended uncertainty relation or the corresponding modified commutation relations from first-principles approach for generic space-times. To this effect, \emph{the current work has filled this void}!

\textcolor{black}{In the context of EUPs, the choice of the description of the dynamics in $4$-dimensional curved space-time through embedding space (and coordinates) is apt in more ways than one. In general, Darboux's theorem~\cite{darboux1882probleme,Wagner:2021bqz} guarantees the existence of \emph{locally defined} canonically conjugate variables that obey the standard Heisenberg algebra.}
Therefore, the following question arises: Are modified position-momentum algebras simply a coordinate artifact, and how are the effects of the \emph{curvature} of $4$-dimensional space-time captured when using (locally-defined) canonically conjugate variables? We plan to address this in a future work.

Furthermore, within a first approximation (ignoring terms of $\mathcal{O}(\hbar^{2})$), the use of Dirac brackets allows us to completely capture the dynamics of a particle in curved space-time through the modifications to the position-momentum algebra. However, higher-order approximations will involve operator-ordering problems and indeterminacy of the complex constants $c_{i}$ corresponding to the second-class constraints. 

Instead of choosing the observer time to fix the reparametrization gauge, we can select another variable that might make the calculations easier. However, in such situations, we might end up with time-dependent constraints, and these might be handled by expanding the phase space to include time $t$ and a conjugate momentum to it~\cite{Gitman:1990qh,Gavrilov_1993}. 

While most of the effort in the literature is focused on the UV modifications of quantum theory due to gravity, our analysis shows that IR modifications naturally arise when unifying general relativity and quantum theory. The results have direct implications for the entanglement entropy of fields. The entanglement entropy of free quantum fields is divergent, and the divergence is commonly thought of as UV origin~\cite{2010-Das.etal-Book,2013-Braunstein.etal-JHEP}. However, exploiting an inherent scaling symmetry of entanglement entropy, it was shown that the cause of divergence of entanglement entropy
is due to zero modes~\cite{2014-Mallayya.etal-PRD,2020-Chandran.Shanki-PRD,2021-Jain.etal-PRD}. 
Interestingly, this scaling symmetry is present in most well-known systems, 
from the two-coupled harmonic oscillator to quantum scalar fields in spherically symmetric space-time. However, one of the key ingredients that need to be added in the earlier analyses is the identification of the IR cut-off to obtain a finite value of the entanglement entropy. The current study provides a systematic way of introducing such a cut-off in the analysis. This is currently under investigation.

Our result has significant implications for black hole thermodynamics. Because of Hawking radiation, the black hole loses mass through its radiation and becomes hotter. As the mass of the black hole decreases, the infrared corrections will become appreciable and might modify the final stages of black hole evaporation. 

Our rigorous analysis systematically shows that the corrections are intricately connected to the curvature scale in the problem. However, we have yet to identify the exact curvature scalar. From the four --- AdS, dS, Schwarzschild, and Geon --- space-times, we speculate that the corrections are proportional to the square root of the Kretschmann scalar. Therefore, one needs to delve deeper into ways to identify the scalar invariant. This is currently under investigation. 

Our analysis can be extended to spin-1/2 particles as the embedded space-time is flat. However, we must find the matching representation from the modified position-momentum commutation relations. This is currently under investigation.

\begin{acknowledgments}
  The authors thank S. M. Chandran, A. Kempf, V. Nenmeli, and K. Rajeev for the discussions. This work is part of MG's undergraduate project. The work of SS is supported by the MATRICS CRG grant (CRG/2022/002348).
\end{acknowledgments}

\section*{Conflict of Interest Declaration}

The authors declare no conflicts of interest related to this work.

\bibliography{main.bib}

\providecommand{\noopsort}[1]{}\providecommand{\singleletter}[1]{#1}%
\begin{thebibliography}{35}%
\makeatletter
\providecommand \@ifxundefined [1]{%
 \@ifx{#1\undefined}
}%
\providecommand \@ifnum [1]{%
 \ifnum #1\expandafter \@firstoftwo
 \else \expandafter \@secondoftwo
 \fi
}%
\providecommand \@ifx [1]{%
 \ifx #1\expandafter \@firstoftwo
 \else \expandafter \@secondoftwo
 \fi
}%
\providecommand \natexlab [1]{#1}%
\providecommand \enquote  [1]{``#1''}%
\providecommand \bibnamefont  [1]{#1}%
\providecommand \bibfnamefont [1]{#1}%
\providecommand \citenamefont [1]{#1}%
\providecommand \href@noop [0]{\@secondoftwo}%
\providecommand \href [0]{\begingroup \@sanitize@url \@href}%
\providecommand \@href[1]{\@@startlink{#1}\@@href}%
\providecommand \@@href[1]{\endgroup#1\@@endlink}%
\providecommand \@sanitize@url [0]{\catcode `\\12\catcode `\$12\catcode
  `\&12\catcode `\#12\catcode `\^12\catcode `\_12\catcode `\%12\relax}%
\providecommand \@@startlink[1]{}%
\providecommand \@@endlink[0]{}%
\providecommand \url  [0]{\begingroup\@sanitize@url \@url }%
\providecommand \@url [1]{\endgroup\@href {#1}{\urlprefix }}%
\providecommand \urlprefix  [0]{URL }%
\providecommand \Eprint [0]{\href }%
\providecommand \doibase [0]{http://dx.doi.org/}%
\providecommand \selectlanguage [0]{\@gobble}%
\providecommand \bibinfo  [0]{\@secondoftwo}%
\providecommand \bibfield  [0]{\@secondoftwo}%
\providecommand \translation [1]{[#1]}%
\providecommand \BibitemOpen [0]{}%
\providecommand \bibitemStop [0]{}%
\providecommand \bibitemNoStop [0]{.\EOS\space}%
\providecommand \EOS [0]{\spacefactor3000\relax}%
\providecommand \BibitemShut  [1]{\csname bibitem#1\endcsname}%
\let\auto@bib@innerbib\@empty
\bibitem [{\citenamefont {Sakurai}\ and\ \citenamefont
  {Napolitano}(2020)}]{sakurai1995modern}%
  \BibitemOpen
  \bibfield  {author} {\bibinfo {author} {\bibfnamefont {J.~J.}\ \bibnamefont
  {Sakurai}}\ and\ \bibinfo {author} {\bibfnamefont {J.}~\bibnamefont
  {Napolitano}},\ }\href {\doibase 10.1017/9781108587280} {\emph {\bibinfo
  {title} {{Modern Quantum Mechanics}}}}\ (\bibinfo  {publisher} {Cambridge
  University Press},\ \bibinfo {year} {2020})\BibitemShut {NoStop}%
\bibitem [{\citenamefont {Shankar}(2012)}]{shankar2012principles}%
  \BibitemOpen
  \bibfield  {author} {\bibinfo {author} {\bibfnamefont {R.}~\bibnamefont
  {Shankar}},\ }\href {\doibase https://doi.org/10.1007/978-1-4757-0576-8}
  {\emph {\bibinfo {title} {Principles of quantum mechanics}}}\ (\bibinfo
  {publisher} {Springer Science \& Business Media},\ \bibinfo {year}
  {2012})\BibitemShut {NoStop}%
\bibitem [{\citenamefont {Kempf}(1996)}]{kempf1996noncommutative}%
  \BibitemOpen
  \bibfield  {author} {\bibinfo {author} {\bibfnamefont {A.}~\bibnamefont
  {Kempf}},\ }\href {\doibase 10.1103/PhysRevD.54.5174} {\bibfield  {journal}
  {\bibinfo  {journal} {Phys. Rev. D}\ }\textbf {\bibinfo {volume} {54}},\
  \bibinfo {pages} {5174} (\bibinfo {year} {1996})},\ \bibinfo {note}
  {[Erratum: Phys.Rev.D 55, 1114 (1997)]},\ \Eprint
  {http://arxiv.org/abs/hep-th/9602119} {arXiv:hep-th/9602119} \BibitemShut
  {NoStop}%
\bibitem [{\citenamefont {Kempf}(1995)}]{kempf1996path}%
  \BibitemOpen
  \bibfield  {author} {\bibinfo {author} {\bibfnamefont {A.}~\bibnamefont
  {Kempf}},\ }in\ \href@noop {} {\emph {\bibinfo {booktitle} {{Minisemester on
  Quantum Groups and Quantum Spaces}}}}\ (\bibinfo {year} {1995})\ \Eprint
  {http://arxiv.org/abs/hep-th/9603115} {arXiv:hep-th/9603115} \BibitemShut
  {NoStop}%
\bibitem [{\citenamefont {Bolen}\ and\ \citenamefont
  {Cavaglia}(2005)}]{Bolen:2004sq}%
  \BibitemOpen
  \bibfield  {author} {\bibinfo {author} {\bibfnamefont {B.}~\bibnamefont
  {Bolen}}\ and\ \bibinfo {author} {\bibfnamefont {M.}~\bibnamefont
  {Cavaglia}},\ }\href {\doibase 10.1007/s10714-005-0108-x} {\bibfield
  {journal} {\bibinfo  {journal} {Gen. Rel. Grav.}\ }\textbf {\bibinfo {volume}
  {37}},\ \bibinfo {pages} {1255} (\bibinfo {year} {2005})},\ \Eprint
  {http://arxiv.org/abs/gr-qc/0411086} {arXiv:gr-qc/0411086} \BibitemShut
  {NoStop}%
\bibitem [{\citenamefont {Park}(2008)}]{Park:2007az}%
  \BibitemOpen
  \bibfield  {author} {\bibinfo {author} {\bibfnamefont {M.-i.}\ \bibnamefont
  {Park}},\ }\href {\doibase 10.1016/j.physletb.2007.11.090} {\bibfield
  {journal} {\bibinfo  {journal} {Phys. Lett. B}\ }\textbf {\bibinfo {volume}
  {659}},\ \bibinfo {pages} {698} (\bibinfo {year} {2008})},\ \Eprint
  {http://arxiv.org/abs/0709.2307} {arXiv:0709.2307 [hep-th]} \BibitemShut
  {NoStop}%
\bibitem [{\citenamefont {Bambi}\ and\ \citenamefont
  {Urban}(2008)}]{Bambi:2007ty}%
  \BibitemOpen
  \bibfield  {author} {\bibinfo {author} {\bibfnamefont {C.}~\bibnamefont
  {Bambi}}\ and\ \bibinfo {author} {\bibfnamefont {F.~R.}\ \bibnamefont
  {Urban}},\ }\href {\doibase 10.1088/0264-9381/25/9/095006} {\bibfield
  {journal} {\bibinfo  {journal} {Class. Quant. Grav.}\ }\textbf {\bibinfo
  {volume} {25}},\ \bibinfo {pages} {095006} (\bibinfo {year} {2008})},\
  \Eprint {http://arxiv.org/abs/0709.1965} {arXiv:0709.1965 [gr-qc]}
  \BibitemShut {NoStop}%
\bibitem [{\citenamefont {Costa~Filho}\ \emph {et~al.}(2016)\citenamefont
  {Costa~Filho}, \citenamefont {Braga}, \citenamefont {Lira},\ and\
  \citenamefont {Andrade~Jr}}]{costa2016extended}%
  \BibitemOpen
  \bibfield  {author} {\bibinfo {author} {\bibfnamefont {R.~N.}\ \bibnamefont
  {Costa~Filho}}, \bibinfo {author} {\bibfnamefont {J.~P.}\ \bibnamefont
  {Braga}}, \bibinfo {author} {\bibfnamefont {J.~H.}\ \bibnamefont {Lira}}, \
  and\ \bibinfo {author} {\bibfnamefont {J.~S.}\ \bibnamefont {Andrade~Jr}},\
  }\href {\doibase https://doi.org/10.1016/j.physletb.2016.02.035} {\bibfield
  {journal} {\bibinfo  {journal} {Phys. Lett. B}\ }\textbf {\bibinfo {volume}
  {755}},\ \bibinfo {pages} {367} (\bibinfo {year} {2016})}\BibitemShut
  {NoStop}%
\bibitem [{\citenamefont {Wagner}(2022)}]{wagner2022relativistic}%
  \BibitemOpen
  \bibfield  {author} {\bibinfo {author} {\bibfnamefont {F.}~\bibnamefont
  {Wagner}},\ }\href {\doibase https://doi.org/10.1103/PhysRevD.105.025005}
  {\bibfield  {journal} {\bibinfo  {journal} {Phys.\ Rev.\ D}\ }\textbf
  {\bibinfo {volume} {105}},\ \bibinfo {pages} {025005} (\bibinfo {year}
  {2022})}\BibitemShut {NoStop}%
\bibitem [{\citenamefont {Petruzziello}\ and\ \citenamefont
  {Wagner}(2021)}]{petruzziello2021gravitationally}%
  \BibitemOpen
  \bibfield  {author} {\bibinfo {author} {\bibfnamefont {L.}~\bibnamefont
  {Petruzziello}}\ and\ \bibinfo {author} {\bibfnamefont {F.}~\bibnamefont
  {Wagner}},\ }\href {\doibase 10.1103/PhysRevD.103.104061} {\bibfield
  {journal} {\bibinfo  {journal} {Phys. Rev. D}\ }\textbf {\bibinfo {volume}
  {103}},\ \bibinfo {pages} {104061} (\bibinfo {year} {2021})},\ \Eprint
  {http://arxiv.org/abs/2101.05552} {arXiv:2101.05552 [gr-qc]} \BibitemShut
  {NoStop}%
\bibitem [{\citenamefont {Dirac}(1950)}]{dirac1950generalized}%
  \BibitemOpen
  \bibfield  {author} {\bibinfo {author} {\bibfnamefont {P.~A.~M.}\
  \bibnamefont {Dirac}},\ }\href {\doibase
  https://doi.org/10.4153/CJM-1950-012-1} {\bibfield  {journal} {\bibinfo
  {journal} {Canadian journal of mathematics}\ }\textbf {\bibinfo {volume}
  {2}},\ \bibinfo {pages} {129} (\bibinfo {year} {1950})}\BibitemShut {NoStop}%
\bibitem [{\citenamefont {Nash}(1956)}]{Nash1956-mv}%
  \BibitemOpen
  \bibfield  {author} {\bibinfo {author} {\bibfnamefont {J.}~\bibnamefont
  {Nash}},\ }\href@noop {} {\bibfield  {journal} {\bibinfo  {journal} {Ann.
  Math.}\ }\textbf {\bibinfo {volume} {63}},\ \bibinfo {pages} {20} (\bibinfo
  {year} {1956})}\BibitemShut {NoStop}%
\bibitem [{\citenamefont {Lidsey}\ \emph {et~al.}(1997)\citenamefont {Lidsey},
  \citenamefont {Romero}, \citenamefont {Tavakol},\ and\ \citenamefont
  {Rippl}}]{lidsey1997applications}%
  \BibitemOpen
  \bibfield  {author} {\bibinfo {author} {\bibfnamefont {J.~E.}\ \bibnamefont
  {Lidsey}}, \bibinfo {author} {\bibfnamefont {C.}~\bibnamefont {Romero}},
  \bibinfo {author} {\bibfnamefont {R.}~\bibnamefont {Tavakol}}, \ and\
  \bibinfo {author} {\bibfnamefont {S.}~\bibnamefont {Rippl}},\ }\href
  {\doibase https://doi.org/10.1088/0264-9381/14/4/005} {\bibfield  {journal}
  {\bibinfo  {journal} {Class. Quant. Grav.}\ }\textbf {\bibinfo {volume}
  {14}},\ \bibinfo {pages} {865} (\bibinfo {year} {1997})}\BibitemShut
  {NoStop}%
\bibitem [{\citenamefont {Friedman}(1965)}]{friedman1965isometric}%
  \BibitemOpen
  \bibfield  {author} {\bibinfo {author} {\bibfnamefont {A.}~\bibnamefont
  {Friedman}},\ }\href {\doibase https://doi.org/10.1103/RevModPhys.37.201.2}
  {\bibfield  {journal} {\bibinfo  {journal} {Rev. Mod. Phys.}\ }\textbf
  {\bibinfo {volume} {37}},\ \bibinfo {pages} {201} (\bibinfo {year}
  {1965})}\BibitemShut {NoStop}%
\bibitem [{\citenamefont {Wald}(2009)}]{wald2009formulation}%
  \BibitemOpen
  \bibfield  {author} {\bibinfo {author} {\bibfnamefont {R.}~\bibnamefont
  {Wald}},\ }\href {\doibase https://doi.org/10.48550/arXiv.0907.0416}
  {\bibfield  {journal} {\bibinfo  {journal} {arXiv preprint arXiv:0907.0416}\
  } (\bibinfo {year} {2009}),\
  https://doi.org/10.48550/arXiv.0907.0416}\BibitemShut {NoStop}%
\bibitem [{\citenamefont {Weinberg}(1995)}]{weinberg1995quantum}%
  \BibitemOpen
  \bibfield  {author} {\bibinfo {author} {\bibfnamefont {S.}~\bibnamefont
  {Weinberg}},\ }\href {\doibase https://doi.org/10.1017/CBO9781139644167}
  {\emph {\bibinfo {title} {The quantum theory of fields}}},\ Vol.~\bibinfo
  {volume} {2}\ (\bibinfo  {publisher} {Cambridge university press},\ \bibinfo
  {year} {1995})\BibitemShut {NoStop}%
\bibitem [{\citenamefont {Wigner}(1939)}]{wigner1939unitary}%
  \BibitemOpen
  \bibfield  {author} {\bibinfo {author} {\bibfnamefont {E.}~\bibnamefont
  {Wigner}},\ }\href {\doibase https://doi.org/10.2307/1968551} {\bibfield
  {journal} {\bibinfo  {journal} {Annals of mathematics}\ }\textbf {\bibinfo
  {volume} {40}},\ \bibinfo {pages} {149} (\bibinfo {year} {1939})}\BibitemShut
  {NoStop}%
\bibitem [{\citenamefont {Moylan}(1983)}]{moylan1983unitary}%
  \BibitemOpen
  \bibfield  {author} {\bibinfo {author} {\bibfnamefont {P.}~\bibnamefont
  {Moylan}},\ }\href {\doibase https://doi.org/10.1063/1.526795} {\bibfield
  {journal} {\bibinfo  {journal} {J. of Math. Phys.}\ }\textbf {\bibinfo
  {volume} {24}},\ \bibinfo {pages} {2706} (\bibinfo {year}
  {1983})}\BibitemShut {NoStop}%
\bibitem [{\citenamefont {Thomas}(1941)}]{thomas1941unitary}%
  \BibitemOpen
  \bibfield  {author} {\bibinfo {author} {\bibfnamefont {L.}~\bibnamefont
  {Thomas}},\ }\href {\doibase https://doi.org/10.2307/1968990} {\bibfield
  {journal} {\bibinfo  {journal} {Annals of mathematics}\ ,\ \bibinfo {pages}
  {113}} (\bibinfo {year} {1941})}\BibitemShut {NoStop}%
\bibitem [{\citenamefont {{Wagner}}\ and\ \citenamefont
  {{Guthrie}}(2021)}]{2021-Wagner.Guthrie-Arx}%
  \BibitemOpen
  \bibfield  {author} {\bibinfo {author} {\bibfnamefont {G.}~\bibnamefont
  {{Wagner}}}\ and\ \bibinfo {author} {\bibfnamefont {M.~W.}\ \bibnamefont
  {{Guthrie}}},\ }\href {\doibase 10.48550/arXiv.2108.07786} {\bibfield
  {journal} {\bibinfo  {journal} {arXiv e-prints}\ ,\ \bibinfo {eid}
  {arXiv:2108.07786}} (\bibinfo {year} {2021})},\ \Eprint
  {http://arxiv.org/abs/2108.07786} {arXiv:2108.07786} \BibitemShut {NoStop}%
\bibitem [{\citenamefont {Fulop}\ \emph {et~al.}(1998)\citenamefont {Fulop},
  \citenamefont {Gitman},\ and\ \citenamefont
  {Tyutin}}]{fulop1998reparametrization}%
  \BibitemOpen
  \bibfield  {author} {\bibinfo {author} {\bibfnamefont {G.}~\bibnamefont
  {Fulop}}, \bibinfo {author} {\bibfnamefont {D.~M.}\ \bibnamefont {Gitman}}, \
  and\ \bibinfo {author} {\bibfnamefont {I.~V.}\ \bibnamefont {Tyutin}},\
  }\href {\doibase https://doi.org/10.48550/arXiv.hep-th/9805040} {\enquote
  {\bibinfo {title} {Reparametrization invariance as gauge symmetry},}\ }
  (\bibinfo {year} {1998}),\ \Eprint {http://arxiv.org/abs/hep-th/9805040}
  {arXiv:hep-th/9805040 [hep-th]} \BibitemShut {NoStop}%
\bibitem [{\citenamefont {Gibbons}(2000)}]{10.1007/3-540-46671-1_5}%
  \BibitemOpen
  \bibfield  {author} {\bibinfo {author} {\bibfnamefont {G.~W.}\ \bibnamefont
  {Gibbons}},\ }in\ \href {\doibase https://doi.org/10.48550/arXiv.1110.1206}
  {\emph {\bibinfo {booktitle} {Mathematical and Quantum Aspects of Relativity
  and Cosmology}}},\ \bibinfo {editor} {edited by\ \bibinfo {editor}
  {\bibfnamefont {S.}~\bibnamefont {Cotsakis}}\ and\ \bibinfo {editor}
  {\bibfnamefont {G.~W.}\ \bibnamefont {Gibbons}}}\ (\bibinfo  {publisher}
  {Springer Berlin Heidelberg},\ \bibinfo {address} {Berlin, Heidelberg},\
  \bibinfo {year} {2000})\ pp.\ \bibinfo {pages} {102--142}\BibitemShut
  {NoStop}%
\bibitem [{\citenamefont {Mignemi}(2010)}]{mignemi2010extended}%
  \BibitemOpen
  \bibfield  {author} {\bibinfo {author} {\bibfnamefont {S.}~\bibnamefont
  {Mignemi}},\ }\href {\doibase https://doi.org/10.1142/S0217732310033426}
  {\bibfield  {journal} {\bibinfo  {journal} {Mod.\ Phys. Letts. A}\ }\textbf
  {\bibinfo {volume} {25}},\ \bibinfo {pages} {1697} (\bibinfo {year}
  {2010})}\BibitemShut {NoStop}%
\bibitem [{\citenamefont {Mallett}\ and\ \citenamefont
  {Fleming}(1973)}]{mallett1973position}%
  \BibitemOpen
  \bibfield  {author} {\bibinfo {author} {\bibfnamefont {R.}~\bibnamefont
  {Mallett}}\ and\ \bibinfo {author} {\bibfnamefont {G.}~\bibnamefont
  {Fleming}},\ }\href {\doibase https://doi.org/10.1063/1.1666170} {\bibfield
  {journal} {\bibinfo  {journal} {J. Math. Phys.}\ }\textbf {\bibinfo {volume}
  {14}},\ \bibinfo {pages} {45} (\bibinfo {year} {1973})}\BibitemShut {NoStop}%
\bibitem [{\citenamefont {Rosen}(1965)}]{rosen1965embedding}%
  \BibitemOpen
  \bibfield  {author} {\bibinfo {author} {\bibfnamefont {J.}~\bibnamefont
  {Rosen}},\ }\href {\doibase https://doi.org/10.1103/RevModPhys.37.204}
  {\bibfield  {journal} {\bibinfo  {journal} {Rev. Mod. Phys.}\ }\textbf
  {\bibinfo {volume} {37}},\ \bibinfo {pages} {204} (\bibinfo {year}
  {1965})}\BibitemShut {NoStop}%
\bibitem [{\citenamefont {Shankaranarayanan}\ and\ \citenamefont
  {Johnson}(2022)}]{2022-Johnson.Shanki-GRG}%
  \BibitemOpen
  \bibfield  {author} {\bibinfo {author} {\bibfnamefont {S.}~\bibnamefont
  {Shankaranarayanan}}\ and\ \bibinfo {author} {\bibfnamefont {J.~P.}\
  \bibnamefont {Johnson}},\ }\href {\doibase 10.1007/s10714-022-02927-2}
  {\bibfield  {journal} {\bibinfo  {journal} {Gen. Rel. Grav.}\ }\textbf
  {\bibinfo {volume} {54}},\ \bibinfo {pages} {44} (\bibinfo {year} {2022})},\
  \Eprint {http://arxiv.org/abs/2204.06533} {arXiv:2204.06533 [gr-qc]}
  \BibitemShut {NoStop}%
\bibitem [{\citenamefont {Darboux}(1882)}]{darboux1882probleme}%
  \BibitemOpen
  \bibfield  {author} {\bibinfo {author} {\bibfnamefont {G.}~\bibnamefont
  {Darboux}},\ }\href {http://www.numdam.org/item/BSMA_1882_2_6_1_14_1/}
  {\bibfield  {journal} {\bibinfo  {journal} {Bulletin des sciences
  math{\'e}matiques et astronomiques}\ }\textbf {\bibinfo {volume} {6}},\
  \bibinfo {pages} {14} (\bibinfo {year} {1882})}\BibitemShut {NoStop}%
\bibitem [{\citenamefont {Wagner}(2021)}]{Wagner:2021bqz}%
  \BibitemOpen
  \bibfield  {author} {\bibinfo {author} {\bibfnamefont {F.}~\bibnamefont
  {Wagner}},\ }\href {\doibase 10.1103/PhysRevD.104.126010} {\bibfield
  {journal} {\bibinfo  {journal} {Phys. Rev. D}\ }\textbf {\bibinfo {volume}
  {104}},\ \bibinfo {pages} {126010} (\bibinfo {year} {2021})},\ \Eprint
  {http://arxiv.org/abs/2110.11067} {arXiv:2110.11067 [gr-qc]} \BibitemShut
  {NoStop}%
\bibitem [{\citenamefont {Gitman}\ and\ \citenamefont
  {Tyutin}(1990)}]{Gitman:1990qh}%
  \BibitemOpen
  \bibfield  {author} {\bibinfo {author} {\bibfnamefont {D.~M.}\ \bibnamefont
  {Gitman}}\ and\ \bibinfo {author} {\bibfnamefont {I.~V.}\ \bibnamefont
  {Tyutin}},\ }\href {\doibase https://doi.org/10.1007/978-3-642-83938-2}
  {\emph {\bibinfo {title} {{Quantization of Fields with Constraints}}}},\
  Springer Series in Nuclear and Particle Physics\ (\bibinfo  {publisher}
  {Springer},\ \bibinfo {address} {Berlin, Germany},\ \bibinfo {year}
  {1990})\BibitemShut {NoStop}%
\bibitem [{\citenamefont {Gavrilov}\ and\ \citenamefont
  {Gitman}(1993)}]{Gavrilov_1993}%
  \BibitemOpen
  \bibfield  {author} {\bibinfo {author} {\bibfnamefont {S.~P.}\ \bibnamefont
  {Gavrilov}}\ and\ \bibinfo {author} {\bibfnamefont {D.~M.}\ \bibnamefont
  {Gitman}},\ }\href {\doibase 10.1088/0264-9381/10/1/008} {\bibfield
  {journal} {\bibinfo  {journal} {Class. Quant. Grav.}\ }\textbf {\bibinfo
  {volume} {10}},\ \bibinfo {pages} {57} (\bibinfo {year} {1993})}\BibitemShut
  {NoStop}%
\bibitem [{\citenamefont {Das}\ \emph {et~al.}(2010)\citenamefont {Das},
  \citenamefont {Shankaranarayanan},\ and\ \citenamefont
  {Sur}}]{2010-Das.etal-Book}%
  \BibitemOpen
  \bibfield  {author} {\bibinfo {author} {\bibfnamefont {S.}~\bibnamefont
  {Das}}, \bibinfo {author} {\bibfnamefont {S.}~\bibnamefont
  {Shankaranarayanan}}, \ and\ \bibinfo {author} {\bibfnamefont
  {S.}~\bibnamefont {Sur}},\ }\enquote {\bibinfo {title} {Black hole entropy
  from entanglement: A review},}\ \ (\bibinfo  {publisher} {Nova Publishers},\
  \bibinfo {year} {2010})\ Chap.~\bibinfo {chapter} {6},\ \Eprint
  {http://arxiv.org/abs/0806.0402} {0806.0402} \BibitemShut {NoStop}%
\bibitem [{\citenamefont {Braunstein}\ \emph {et~al.}(2013)\citenamefont
  {Braunstein}, \citenamefont {Das},\ and\ \citenamefont
  {Shankaranarayanan}}]{2013-Braunstein.etal-JHEP}%
  \BibitemOpen
  \bibfield  {author} {\bibinfo {author} {\bibfnamefont {S.~L.}\ \bibnamefont
  {Braunstein}}, \bibinfo {author} {\bibfnamefont {S.}~\bibnamefont {Das}}, \
  and\ \bibinfo {author} {\bibfnamefont {S.}~\bibnamefont
  {Shankaranarayanan}},\ }\href {\doibase 10.1007/JHEP07(2013)130} {\bibfield
  {journal} {\bibinfo  {journal} {JHEP}\ }\textbf {\bibinfo {volume} {07}},\
  \bibinfo {pages} {130} (\bibinfo {year} {2013})},\ \Eprint
  {http://arxiv.org/abs/1110.1239} {arXiv:1110.1239 [hep-th]} \BibitemShut
  {NoStop}%
\bibitem [{\citenamefont {Mallayya}\ \emph {et~al.}(2014)\citenamefont
  {Mallayya}, \citenamefont {Tibrewala}, \citenamefont {Shankaranarayanan},\
  and\ \citenamefont {Padmanabhan}}]{2014-Mallayya.etal-PRD}%
  \BibitemOpen
  \bibfield  {author} {\bibinfo {author} {\bibfnamefont {K.}~\bibnamefont
  {Mallayya}}, \bibinfo {author} {\bibfnamefont {R.}~\bibnamefont {Tibrewala}},
  \bibinfo {author} {\bibfnamefont {S.}~\bibnamefont {Shankaranarayanan}}, \
  and\ \bibinfo {author} {\bibfnamefont {T.}~\bibnamefont {Padmanabhan}},\
  }\href {\doibase 10.1103/PhysRevD.90.044058} {\bibfield  {journal} {\bibinfo
  {journal} {Phys. Rev. D}\ }\textbf {\bibinfo {volume} {90}},\ \bibinfo
  {pages} {044058} (\bibinfo {year} {2014})},\ \Eprint
  {http://arxiv.org/abs/1404.2079} {arXiv:1404.2079 [hep-th]} \BibitemShut
  {NoStop}%
\bibitem [{\citenamefont {Chandran}\ and\ \citenamefont
  {Shankaranarayanan}(2020)}]{2020-Chandran.Shanki-PRD}%
  \BibitemOpen
  \bibfield  {author} {\bibinfo {author} {\bibfnamefont {S.~M.}\ \bibnamefont
  {Chandran}}\ and\ \bibinfo {author} {\bibfnamefont {S.}~\bibnamefont
  {Shankaranarayanan}},\ }\href {\doibase 10.1103/PhysRevD.102.125025}
  {\bibfield  {journal} {\bibinfo  {journal} {Phys. Rev. D}\ }\textbf {\bibinfo
  {volume} {102}},\ \bibinfo {pages} {125025} (\bibinfo {year} {2020})},\
  \Eprint {http://arxiv.org/abs/2010.03418} {arXiv:2010.03418 [gr-qc]}
  \BibitemShut {NoStop}%
\bibitem [{\citenamefont {Jain}\ \emph {et~al.}(2021)\citenamefont {Jain},
  \citenamefont {Chandran},\ and\ \citenamefont
  {Shankaranarayanan}}]{2021-Jain.etal-PRD}%
  \BibitemOpen
  \bibfield  {author} {\bibinfo {author} {\bibfnamefont {P.}~\bibnamefont
  {Jain}}, \bibinfo {author} {\bibfnamefont {S.~M.}\ \bibnamefont {Chandran}},
  \ and\ \bibinfo {author} {\bibfnamefont {S.}~\bibnamefont
  {Shankaranarayanan}},\ }\href {\doibase 10.1103/PhysRevD.103.125008}
  {\bibfield  {journal} {\bibinfo  {journal} {Phys. Rev. D}\ }\textbf {\bibinfo
  {volume} {103}},\ \bibinfo {pages} {125008} (\bibinfo {year} {2021})},\
  \Eprint {http://arxiv.org/abs/2103.01772} {arXiv:2103.01772 [hep-th]}
  \BibitemShut {NoStop}%
\end{thebibliography}%
\end{document}